\documentclass[%
 reprint,
 amsmath,amssymb,
 aps,
]{revtex4-2}

\usepackage{stmaryrd}
\usepackage{bm}
\usepackage{float}
\usepackage{color}
\usepackage{braket}
\usepackage{amsmath}
\usepackage{algorithmic}
\usepackage{algorithm}
\usepackage[pdftex]{graphicx}
\usepackage[pdftex]{overpic}

\newcommand{\er}[1]{\textcolor{yellow}{}}
\renewcommand{\S}{\mathcal{S}}

\newcommand{\E}{\mathcal{E}}

\newcommand{\J}{\mathcal{J}}
\renewcommand{\L}{\mathcal{L}}

\renewcommand{\ss}{{\bm s}}

\renewcommand{\tt}{{\bm t}}

\newcommand{\argmax}{\mathop{\rm arg~max}\limits}

\newcommand{\inn}{\text{in}}
\newcommand{\out}{\text{out}}
\newcommand{\bkt}[1]{\llbracket{#1}\rrbracket}

\usepackage{bbm}
\usepackage{multirow}
\begin{document}

\preprint{APS/123-QED}

\title{Approximate level-by-level maximum-likelihood decoding based on the Chase algorithm for high-rate concatenated stabilizer codes}

\author{Takeshi Kakizaki}

 \affiliation{%
 National Institute of Advanced Industrial Science and Technology, Koto-ku, Tokyo 135-0064, Japan}%

\date{\today}

\begin{abstract}
Fault-tolerant quantum computation (FTQC) is expected to address a wide range of computational problems. To realize large-scale FTQC, it is essential to encode logical qubits using quantum error-correcting codes. High-rate concatenated codes have recently attracted attention due to theoretical advances in fault-tolerant protocols with constant-space-overhead and polylogarithmic-time-overhead, as well as practical developments of high-rate many-hypercube codes equipped with a high-performance level-by-level minimum-distance decoder (LMDD). We propose a general, high-performance decoder for high-rate concatenated stabilizer codes that extends LMDD by leveraging the Chase algorithm to generate a suitable set of candidate errors. Our simulation results demonstrate that the proposed decoder outperforms conventional decoders for high-rate concatenated Hamming codes under bit-flip noise.
\end{abstract}

\maketitle

\section{Introduction}
Quantum algorithms are expected to provide computational advantages for important problem classes that are intractable for classical computers \cite{nielsen2010quantum,gottesman2022opportunities}. To enable scalable fault-tolerant quantum computation (FTQC), quantum error correction (QEC) is essential to mitigate the effects of decoherence and hardware imperfections. Stabilizer codes, a general class of quantum error-correcting codes (QECCs), encode $k$ logical qubits into $n$ noisy physical qubits. A decoder for a stabilizer code estimates the logical error from the measured syndromes and a noise model.

Several major families of QECCs have been studied, such as topological codes \cite{KITAEV20032,PhysRevLett.97.180501,bravyi1998quantum,freedman2001projective}, quantum low-density parity-check (qLDPC) codes \cite{gottesman2013fault,tamiya2025fault,k4cm-pp9p}, and concatenated codes \cite{knill2005quantum,yamasaki2024time}. Topological codes have been among the most promising architectures in recent years, owing to their locality and low stabilizer weight. Experimental demonstrations have been reported on superconducting platforms \cite{barends2014superconducting,google2023suppressing,alphaqubit,sivak2025reinforcement,google2025quantum}. qLDPC codes are attractive due to their low space overhead compared with surface codes, while generally featuring nonlocality \cite{bravyi2010tradeoffs}. Such nonlocality has become increasingly feasible in recent years, as demonstrated by experimental advances using ion traps \cite{ryan2021realization,ryan2024high,paetznick2024demonstration} and neutral-atom platforms \cite{evered2023high,bluvstein2024logical,xu2024constant}.

\begin{figure*}[ht!]
  \centering
  \begin{overpic}[width=12cm]{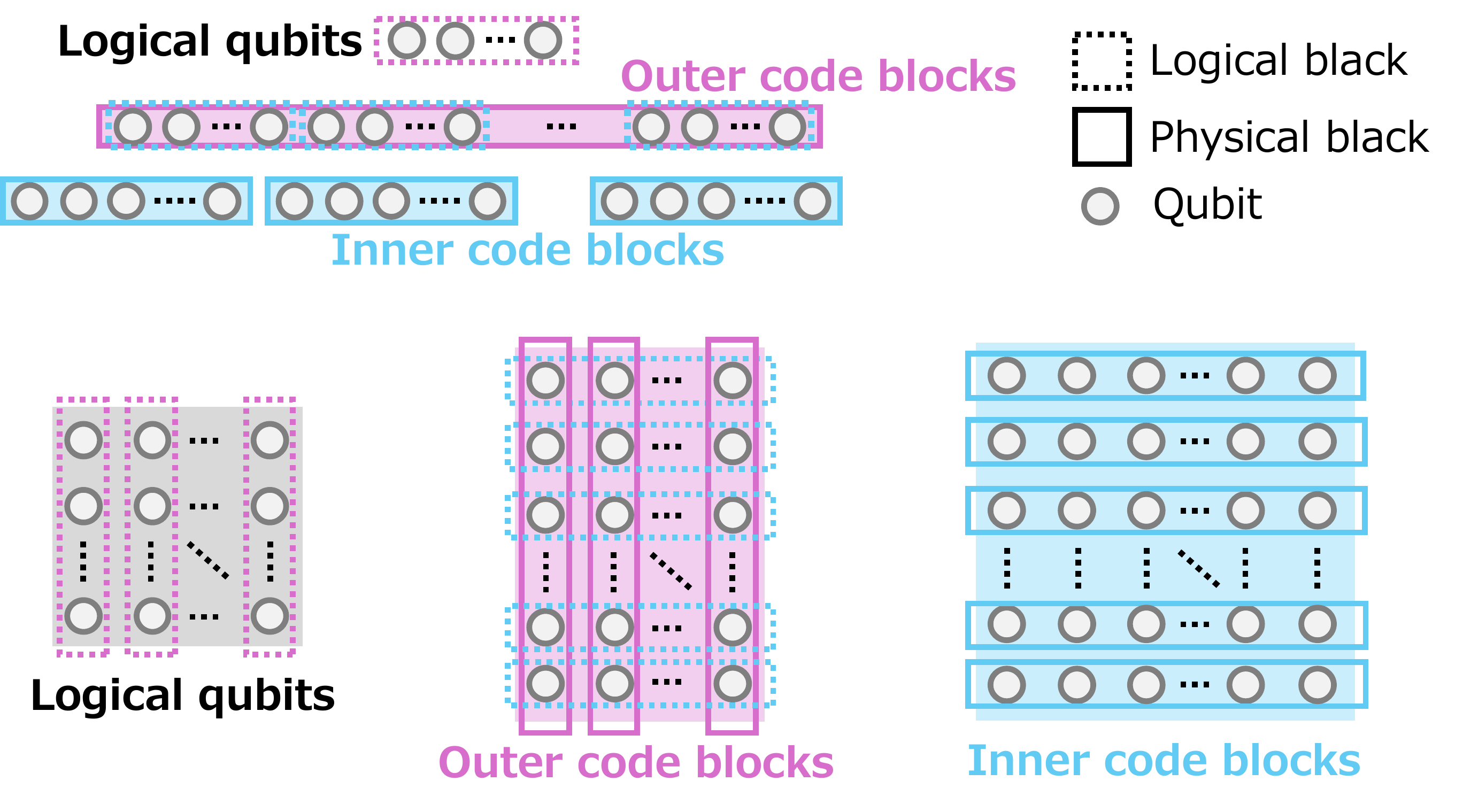}
    \put(-5,53){(a)}
    \put(-5,33){(b)}
    \put(94,29){$E_{[1,:]}$}
    \put(94,24.5){$E_{[2,:]}$}
    \put(94,20){$E_{[3,:]}$}
    \put(94,12){$E_{[n^{\out}-1,:]}$}
    \put(94,8){$E_{[n^{\out},:]}$}
    \put(35,34){$L_{[:,1]}$}
    \put(49,34){$L_{[:,k^{\inn}]}$}
    \put(54,29){$L_{[1,:]}$}
    \put(54,8){$L_{[n^{\out},:]}$}
    \end{overpic}
    \caption{Schematic of (a) low-rate concatenated codes and (b) high-rate concatenated codes.}
    \label{fig:concatenated_code}
\end{figure*}
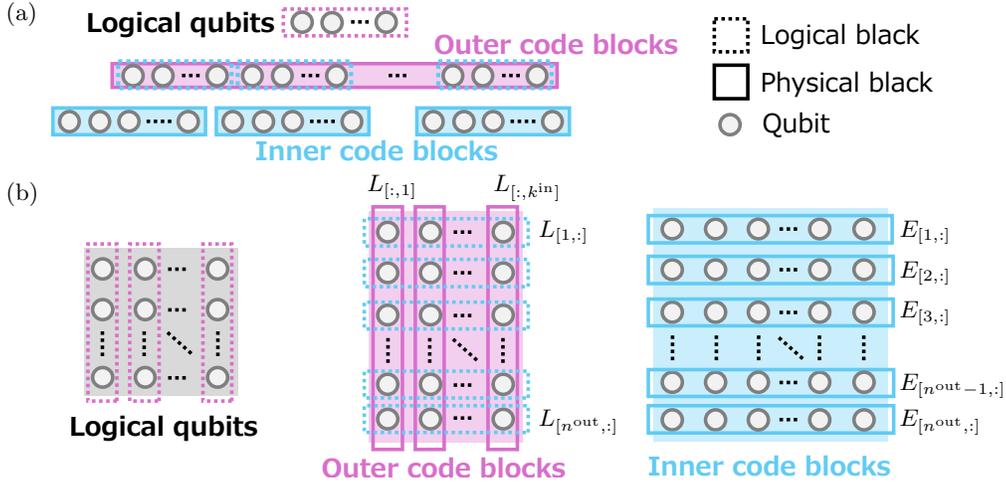

In recent years, concatenated codes have attracted renewed attention due to the development of low-overhead fault-tolerant (FT) protocols based on high-rate concatenated Hamming codes, which achieve constant-space-overhead and quasi-polylogarithmic time overhead \cite{yamasaki2024time,yoshida2025concatenate}, as well as high-rate many-hypercube (MHC) code families \cite{goto2024high,nakai2025subsystem,goto2025optimized}. High-rate concatenated codes achieve a large number of logical qubits by parallelizing and interleaving the inner and outer codes. This is in contrast to traditional low-rate concatenated codes, as shown in Fig. \ref{fig:concatenated_code}, which have been extensively studied in the context of the threshold theorem, optimal decoders \cite{poulin2006optimal}, and FT protocols \cite{knill2005quantum}, and whose rates decrease rapidly with increasing concatenation level $r$.

Achieving high noise tolerance requires not only a well-designed QECC but also a high-performance decoder capable of approaching its high performance. A major challenge for high-rate concatenated codes lies in the design of an effective decoder. As in low-rate concatenated codes, a simple hard-decision decoder (HDD) can be readily constructed, but its decoding performance is severely limited. In high-rate concatenated codes, a large number of logical qubits are encoded across multiple blocks through parallelization and interleaving, a structure that can propagate physical errors to higher concatenation levels. A symbol-by-symbol maximum a posteriori (symbol-MAP) decoder improves decoding performance over HDD by estimating each logical qubit independently. However, it remains suboptimal because it neglects correlations among multiple logical qubits. For small minimum-distance codes such as many hypercube codes, a level-by-level minimum-distance decoder (LMDD) has been proposed, which identifies a minimum-distance error at each level and achieves performance exceeding that of symbol-MAP decoders \cite{goto2024high}.

In this paper, we propose a level-by-level maximum-likelihood decoder based on the Chase algorithm (LMLD-CA), which extends the conventional LMDD developed for MHCs to high-rate concatenated codes consisting of inner and outer stabilizer codes with relatively small minimum distance. The proposed LMLD-CA estimates the most likely logical error by efficiently searching a subset of physical-error candidates and calculating their probabilities using the Chase algorithm. We demonstrate that LMLD-CA achieves superior decoding performance compared with conventional HDD and symbol-MAP decoders when applied to concatenated Hamming codes. As a result, LMLD-CA provides a versatile and efficient decoding framework for high-rate concatenated codes. Recent work has demonstrated that bidirectional HDD can outperform conventional HDD by using higher-level syndromes to reassign the errors~\cite{zhang2026bidirectional}. In contrast, our approach adopts a list-decoding-type decoding strategy, in which multiple candidate errors and their associated probabilities are recursively generated and evaluated.

\section{Conventional Decoder for high-rate concatenated codes}
In this section, we describe high-rate concatenated codes and their conventional decoders. We consider high-rate concatenated codes with $\bkt{n^\inn,k^\inn,d^\inn}$ inner codes and $\bkt{n^\out,k^\out,d^\out}$ outer codes, in which multiple inner code blocks are encoded in parallel, as shown in Fig.~\ref{fig:concatenated_code} (b). The physical error $E$ is decomposed as
\begin{align}
    E = E_{[1,:]} E_{[2,:]} \cdots E_{[n^\out,:]}.
\end{align}
Here, $E_{[i,:]}$ denotes the inner-code physical-error block and is defined as
\begin{align}
    E_{[i,:]} = E_{[i,1]} E_{[i,2]} \cdots E_{[i,n^\inn]}.
\end{align}
The corresponding inner-code logical error block is given by
\begin{align}
    L_{[i,:]} = L_{[i,1]} L_{[i,2]} \cdots L_{[i,k^\inn]},
\end{align}
for each inner-code block index $1 \leq i \leq n^{\out}$. The inner-code logical errors are treated as outer-code physical errors and written as
\begin{align}
    L_{[:,j]} = L_{[1,j]} L_{[2,j]} \cdots L_{[n^\out,j]}
\end{align}
for each outer-code block index $1 \leq j \leq k^{\inn}$.

\subsection{HDD}
The recovery operators of the inner and outer codes are defined as $T^\inn(\ss_i^\inn)$ and $T^\out(\ss_j^\out)$, respectively, where the inner- and outer-code syndromes are given by $\ss_i^\inn=\S^\inn(E_{[i,:]})$ and $\ss_j^\out=\S^\out(E_{[:,j]})$. The inner-code HDD estimates the inner-code physical errors $\hat{E}_{[i,:]}=T^\inn(\ss_i^\inn)$ from the inner-code syndrome $\ss^\inn_i$ for each $1\leq i \leq n^\out$. The inner-code HDD succeeds if $\hat{E}_{[i,:]}E_{[i,:]}\in\S^\inn$, and fails by inducing an inner-code logical error when $L_{[i,:]}=\hat{E}_{[i,:]}E_{[i,:]}\in\L$, where $\L$ denotes the set of logical errors. The outer-code HDD then corrects the remaining outer-code physical errors $L_{[:,j]}$ based on the syndrome $\ss_j^\out$ for each $1\leq j \leq k^\inn$.

\subsection{Symbol-by-symbol MAP decoder}
In this section, we describe a symbol-MAP decoder, which uses inner-code logical error probabilities to estimate the outer-code logical error at each level. We define the function $\L(E)$ that returns the logical operator $L$ such that $E = LST(\ss)$. The optimal degenerate quantum maximum likelihood decoder (DQMLD) estimates the most likely logical error according to
\begin{align}
    \hat{L} = \argmax_{L\in \L} P(L|\ss)
    = \argmax_{L\in \L} \sum_{S\in\S} P(LST(\ss)|\ss). \label{eq:dqmld}
\end{align}
The symbol-MAP decoder estimates the marginal probability of each inner-code logical error,
\begin{align}
    P(L_j|\ss) = \sum_{L_1}\cdots\sum_{L_{j-1}}\sum_{L_{j+1}}\cdots\sum_{L_k}P(L_1,L_2,\cdots,L_k|\ss), \label{eq:symbol_by_symbol_dqmld}
\end{align}
for each logical error index $j$. Direct evaluation of the sum over $\sum_{S\in\S}$ and the sums over all logical errors $\sum_{L_1}\cdots\sum_{L_{j-1}}\sum_{L_{j+1}}\cdots\sum_{L_k}$ becomes intractable for large numbers of physical qubits $n$, since the total number of terms grows exponentially in $n$. The Chase algorithm approximates $P(L|\ss)$ with low computational complexity by choosing a subset of physical errors $\E' \subset \E$, under the assumption of a small minimum distance $d$ \cite{11249913}. As a result, the symbol-MAP decoder improves decoding performance compared with HDD by incorporating logical-error probabilities at each level, although its decoding performance is limited by the approximation in Eq. \eqref{eq:symbol_by_symbol_dqmld}. Importantly, the symbol-MAP decoder achieves optimal decoding performance for concatenated codes in which each level encodes only a single logical qubit \cite{poulin2006optimal}.

\subsection{LMDD}
The LMDD provides a high-performance approximation to the MDD for MHCs \cite{goto2024high} and recursively constructs candidate inner-code physical errors. Each candidate consists of the $n^{\out}-1$ minimum-weight outer-code physical-error blocks $L_{[i,:]}$ for all blocks except the $i'$th, together with the remaining block $L_{[i',:]}$, which is inferred from the syndromes and the already determined outer-code physical error blocks. The candidate inner-code physical error with the lowest weight is then obtained by multiplying the stabilizer $X_1X_2\cdots X_n$ (or $Z_1Z_2\cdots Z_n$), thereby providing an estimate of the minimum-weight error. However, when many stabilizers must be considered simultaneously, identifying the minimum-weight set of error blocks becomes challenging. In this case, the outer-code physical error cannot be uniquely inferred from the other blocks, and the undetermined errors cannot be resolved solely from the stabilizer constraints and the already determined errors.

\section{LMLD-CA}
In this section, we describe the LMLD-CA, extending LMDD to general stabilizer codes and is based on the Chase algorithm. LMLD-CA approximates DQMLD by restricting the search to a candidate subset of physical errors $\tilde{\E}\subset\E$ and evaluates their posterior probabilities according to
\begin{align}
    P(E|\ss)\simeq P(E|\ss,\tilde{\E})\propto \mathbbm{1}_{[E\in \tilde{\E}]}\mathbbm{1}_{[\ss=\S(E)]}P(E). \label{eq:qmld}
\end{align}
Here, $\mathbbm{1}[\cdot]$ denotes the indicator function, $\S(E)$ is the syndrome mapping, and $P(E)$ is a prior error probability.

To obtain a subset of physical error, each inner-codes decoder first identifies a subset of inner-code physical errors $\tilde{\E}_i^{\inn}\subset \E_i^{\inn}$ and calculates their conditional probabilities
\begin{align}
    P(E_{[i,:]}|\ss_i^\inn,\tilde{\E}^\inn_i)
    \propto \mathbbm{1}_{[E_{[i,:]}\in \tilde{\E}_i^{\inn}]}
    \mathbbm{1}_{[\ss_i^\inn=\S^\inn(E_{[i,:]})]}
    P(E_{[i,:]}), \label{eq:physical_error_probability}
\end{align}
for each error $E_{[i,:]}\in\tilde{\E}_i^{\inn}$ and for each inner-code block index $1\leq i\leq n^{\out}$. These inner-code candidates need not satisfy the outer-code syndrome condition.

The decoder then generates the list of inner-code logical errors
\begin{align}
    L_{[i,:]}^{(1)}, L_{[i,:]}^{(2)}, \cdots, L_{[i,:]}^{(|\tilde{\E}_i^\inn|)},
\end{align}
sorted in descending order of probability, as $P(L_{[i,:]}^{(\alpha)}|\ss^\inn_i,\tilde{\E}^\inn_i) \geq P(L_{[i,:]}^{(\beta)}|\ss^\inn_i,\tilde{\E}^\inn_i)$ for any $\alpha<\beta$. The reliability $\gamma_i$ is defined as
\begin{align}
    \gamma_i \triangleq \log \frac{P(L_{[i,:]}^{(1)}|\ss_i^\inn,\tilde{\E}_i^\inn)}{P(L_{[i,:]}^{(2)}|\ss_i^\inn,\tilde{\E}_i^\inn)}\geq 0,
\end{align}
where $L_{[i,:]}^{(1)}$ and $L_{[i,:]}^{(2)}$ are the most- and second-most-likely inner-code logical errors, respectively.

The decoder then constructs test patterns (TPs) $L^{(\ell)}$ for each $1\leq \ell \leq \Lambda$. These TPs are generated from the $D$-list of logical error candidates accoss $M$ selected block indices. Let $\J := \{\sigma(1),\sigma(2),\cdots,\sigma(M)\}$, where $\sigma$ is a permutation of $\{1,2,\cdots,n^\out\}$ such that $\gamma_{\sigma(1)} \leq \gamma_{\sigma(2)} \leq\cdots \leq \gamma_{\sigma(n^\out)}$. For each TP $L^{(\ell)}$, the logical error at block $j$ is set to $\{L_{[:,j]}^{(1)}\}$ if $j \not\in \J$, and from $\{L_{[:,j]}^{(1)}, L_{[:,j]}^{(2)}, \cdots, L_{[:,j]}^{(D)}\}$ otherwise.

The outer decoder calculates candidate outer-code physical errors $\hat{L}_{[:,j]}$ using the HDD, applied to the updated outer-code syndrome $\ss^{\out,(\ell)}_j = \ss_j^{\out} \cdot \S^{\out}(L^{(\ell)})$ for each outer-code block index $j$ and TP index $\ell$. Candidates with identical probabilities may be discarded, which can significantly degrade decoding performance when multiple error candidates are generated and $D$ is small.

The decoder then constructs the estimated inner-code physical error $\hat{E}^{(\ell)}$ for each $1\leq \ell \leq \Lambda$. When $E_{[i,:]}\neq \hat{E}_{[i,:]}^{(\ell)}$, each $\hat{E}_{[i,:]}^{(\ell)}$ is determined as
\begin{align}
    \hat{E}^{(\ell)}_{[i,:]} \triangleq
    \begin{cases}
        \argmax P(E_{[i,:]}^{(\ell)}|\ss_i^\inn) & (E_{[i,:]}^{(\ell)}\in \tilde{\E}^{\inn}_i),\\
        \phi & (E_{[i,:]}^{(\ell)}\not\in \tilde{\E}^{\inn}_i),
    \end{cases}
\end{align}
where $\phi$ denotes a null symbol. Finally, LMLD-CA evaluates Eqs. \eqref{eq:qmld} and \eqref{eq:dqmld} over the restricted subset $\tilde{\E}\triangleq \{\hat{E}^{(\ell)}\}_{1\leq \ell\leq \Lambda}$.

\begin{figure*}[ht!]
  \centering
  \begin{overpic}[width=14cm]{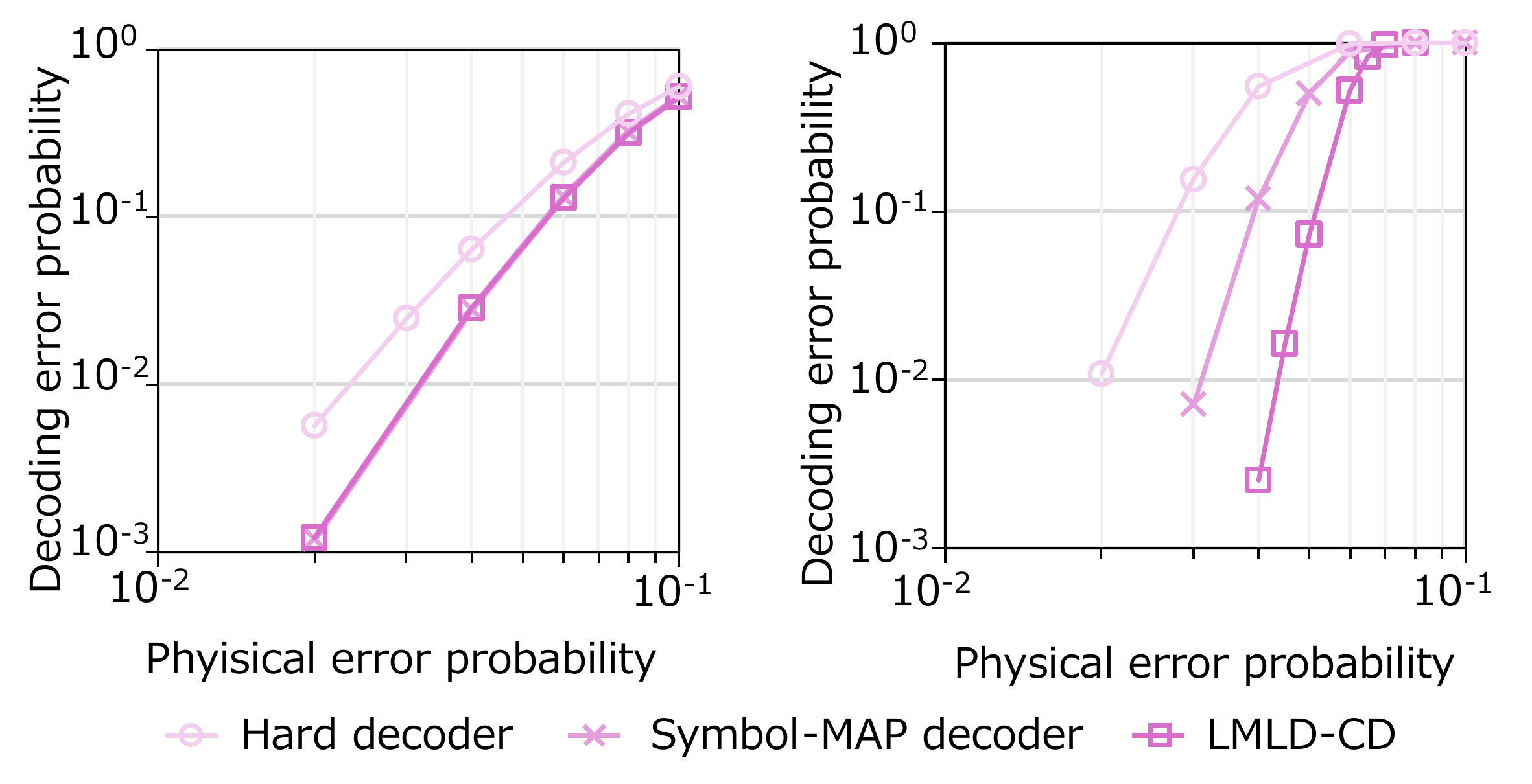}
    \put(-2,50){(a)}
    \put(48,50){(b)}
    \end{overpic}
    \caption{Decoding error probability with bit-flip noise of (a) two- and (b) three-level concatenated Hamming codes.}
    \label{fig:decoding_performance}
\end{figure*}

\section{Decoding performance}\label{sec:decoding_performance}
We present numerical simulations demonstrating that the LMLD-CA can outperform conventional decoders for the two- and three-level concatenated Hamming codes under the bit-flip noise with physical error probability $p$. For the two-level concatenation, Monte Carlo simulations were continued until at least $10^4$, $5\times 10^2$, and $6\times 10^2$ decoding errors were observed for HDD, the symbol-MAP decoder, and LMLD-CA, respectively.
For the three-level concatenation, simulations were continued until at least $10^3$, $10^3$, and $2\times 10^2$ decoding errors were observed, respectively. 

Figure \ref{fig:decoding_performance}(a) compares the HDD, the symbol-MAP (Chase) decoder, and the proposed LMLD-CA for the two-level concatenation. The number of flips $M$ and the list size $D$ are set to $8$ and $2$, respectively, for both the symbol-MAP decoder and LMLD-CA; this setting yields near-saturated decoding performance, consistent with the behavior of the classical decoders based on the Chase algorithm \cite{chase1972class}. The symbol-MAP decoder and LMLD-CA approach optimal decoding performance~\cite{poulin2006optimal}.

In contrast, Figure \ref{fig:decoding_performance}(b) shows the corrsponding results for the three-level concatenated code. Here, LMLD-CA uses a list size of $D=4$. The symbol-MAP decoder suffers from performance degradation due to the approximation in Eq. \eqref{eq:symbol_by_symbol_dqmld}. This is because the $\bkt{15,7,3}$-Hamming code has multiple logical bits. LMLD-CA mitigates this degradation by searching for candidates on a block-by-block basis, thereby significantly improving the decoding performance compared with the symbol-MAP decoder.

\section{Complexity discussion}
LMLD-CA can improve decoding performance compared with the symbol-MAP decoder; however, this improvement comes at the cost of increased decoding complexity, as the list size $D$ grows with an increasing number of concatenation levels $r$. For instance, for $M=8$ and $D=4$, as implemented in Sec.~\ref{sec:decoding_performance}, a straightforward implementation requires evaluating $D^M = 4^8$ candidates, which leads to a substantial computational burden. Analogous to classical decoders based on the Chase algorithm, the number of TPs grows exponentially with the minimum distance $d$, since a large number of TPs is required to identify the most likely error and to accurately approximate its probability. Nevertheless, we demonstrate that LMLD-CA can achieve high decoding performance using a limited number of TPs, thereby providing a practical high-performance decoder for stabilizer codes beyond MHC.

We note that the generation of TP is generally not optimal. The number of simultaneous bit flips can be constrained, since the probability of multiple errors decreases exponentially. There is also room to optimize the list size according to the reliabilities. Alternatively, several well-known low-complexity decoders\cite{ORDGRAND,osd}  may offer lower computational complexity than the Chase algorithm. To assess their practical decoding performance, it is necessary to evaluate not only the properties of the most likely errors they identify, but also the accuracy of the approximated output error probabilities, as is commonly done in the analysis of classical turbo product codes~\cite{10273466}.

\section{Conclusion}
This paper proposed an LMLD-CA for concatenated stabilizer codes that can efficiently approximately estimates the most likely logical error by searching physical error subset based on the Chase algorithm. Our numerical simulation show that the LMLD-CA for concatenated Hamming codes exponentially improve the decoding error probability compared with the convetional decoders.

\begin{acknowledgments}
I would like to thank S. Takizawa for useful discussions about decoding complexity. This work was supported by JSPS KAKENHI Grant Number 25K24401.
\end{acknowledgments}

\bibliography{apssamp}

\end{document}